%% file: verification_of_anticoncentrated_quantum_states_NPJ.tex
\documentclass[english,aps,manuscript,11pt]{revtex4}
\usepackage[T1]{fontenc}
\usepackage[latin9]{inputenc}
\usepackage[letterpaper]{geometry}
\usepackage{amsmath}
\usepackage{amsthm}
\usepackage{graphicx}

\makeatletter
\@ifundefined{textcolor}{}
{%
 \definecolor{BLACK}{gray}{0}
 \definecolor{WHITE}{gray}{1}
 \definecolor{RED}{rgb}{1,0,0}
 \definecolor{GREEN}{rgb}{0,1,0}
 \definecolor{BLUE}{rgb}{0,0,1}
 \definecolor{CYAN}{cmyk}{1,0,0,0}
 \definecolor{MAGENTA}{cmyk}{0,1,0,0}
 \definecolor{YELLOW}{cmyk}{0,0,1,0}
}

\usepackage{color}

\makeatother

\usepackage{babel}
\begin{document}
\input{macros.tex}

\vspace*{2.5in}

This manuscript has been authored by UT-Battelle, LLC, under contract
DE-AC05-00OR22725 with the US Department of Energy (DOE). The US government
retains and the publisher, by accepting the article for publication,
acknowledges that the US government retains a nonexclusive, paid-up,
irrevocable, worldwide license to publish or reproduce the published
form of this manuscript, or allow others to do so, for US government
purposes. DOE will provide public access to these results of federally
sponsored research in accordance with the DOE Public Access Plan (http://energy.gov/downloads/doe-public-access-plan).
\newpage{}
\title{Efficient Verification of Anticoncentrated Quantum States}
\author{Ryan Bennink}
\affiliation{Quantum Computational Science Group, Oak Ridge National Laboratory}
\begin{abstract}
A promising use of quantum computers is to prepare quantum states
that model complex domains, such as correlated electron wavefunctions
or the underlying distribution of a complex dataset. Such states need
to be verified in view of algorithmic approximations and device imperfections.
As quantum computers grow in size, however, verifying the states they
produce becomes increasingly problematic. Relatively efficient methods
have been devised for verifying sparse quantum states, but dense quantum
states have remained costly to verify. Here I present a novel method
for estimating the fidelity $F(\mu,\tau)$ between a preparable quantum
state $\mu$ and a classically specified target state $\tau$, using
simple quantum circuits and on-the-fly classical calculation (or lookup)
of selected amplitudes of $\tau$. Notably, in the targeted regime
the method demonstrates an exponential quantum advantage in sample
efficiency over any classical method. The simplest version of the
method is efficient for anticoncentrated quantum states (including
many states that are hard to simulate classically), with a sample
cost of approximately $4\epsilon^{-2}(1-F)dp_{\text{coll}}$ where
$\epsilon$ is the desired precision of the estimate, $d$ is the
dimension of the Hilbert space in which $\mu$ and $\tau$ reside,
and $p_{\text{coll}}$ is the collision probability of the target
distribution. I also present a more sophisticated version of the method,
which uses any efficiently preparable and well-characterized quantum
state as an importance sampler to further reduce the number of copies
of $\mu$ needed. Though some challenges remain, this work takes a
significant step toward scalable verification of complex states produced
by quantum processors.
\end{abstract}
\maketitle

\section{Introduction}

One of the more promising and imminent applications of quantum computers
is the efficient preparation of quantum states that model complex,
computationally challenging domains. For example, quantum computers
can in principle efficiently simulate states of quantum many-body
systems \citep{Georgescu2014,Tacchino2020} and learn to output states
that model complex data sets \citep{Biamonte2017,Benedetti2019}.
But real quantum processors have imperfections, and the algorithms
themselves often involve approximations whose impact is not fully
understood. Consequently, for the foreseeable future, complex states
produced by quantum processors need to be experimentally verified.
Unfortunately, verifying many-qubit states presents a major challenge
since the number of parameters specifying a quantum state is generally
exponential in the number of qubits. Recently developed quantum processors
are already capable of producing states too large and complex to be
directly verified using available methods \citep{Arute2019}.

Nearly all known methods for efficiently verifying large quantum states
either require the state to have special structure, or require additional
quantum resources. If one has the means to prepare reference copies
of the target state $\tau$, the quantum swap test \citep{Buhrman2001}
can be used to estimate the fidelity of any quantum state $\mu$ with
respect to $\tau$ using $O(1)$ copies of $\tau$. But often, reference
copies of the target state are not available. In such cases, known
structure of the state in question may be used to reduce the number
of measurements needed to characterize it. For example, pure or low-rank
states are described by exponentially fewer parameters than mixed
states (though still exponentially many). Compressive sampling can
be used to efficiently learn quantum states that are sparse in a known
basis \citep{Shabani2011,Jiying2010,Howland2016,Ahn2019}. States
produced by local dynamics can also be efficiently characterized \citep{Bairey2019}.
Notably, it is not necessary to fully characterize an unknown state
in order to verify it. \citep{Flammia2011,daSilva2011} propose the
approach of expressing $F(\mu,\tau)$ as a sum of terms that are sampled
and estimated separately. That approach enables scalable verification
of states in that are concentrated in the Pauli operator basis.

In this Letter I present a novel hybrid (quantum-classical) algorithm,
designated EVAQS, for efficient verification of anticoncentrated quantum
states\footnote{Technically, the concentration of a quantum state depends on the chosen
basis. This work targets states that are \emph{practically} anticoncentrated,
that is, not concentrated in any known and experimentally feasible
measurement basis.}. The algorithm estimates the fidelity $F(\mu,\tau)$ between a preparable
quantum state $\mu$ and a classically-described pure quantum state
$\tau$. The underlying idea of the algorithm is to compare randomly
selected sparse projections or ``snippets'' of the unknown state
$\mu$ against corresponding snippets of the target state $\tau$.
Since the snippets of $\tau$ are small they can be efficiently calculated
and prepared. The fidelity between $\mu$ and $\tau$ is then estimated
as a weighted average of the fidelities of corresponding snippets.
The basic version of EVAQS consists of a simple feed-forward quantum
circuit applied to multiple copies of the prepared state $\mu$ and
two ancilla qubits. The feed-forward part of the circuit involves
an on-the-fly classical calculation (or lookup) of a few randomly-selected
probability amplitudes of $\tau$. (Actually, these amplitudes need
be determined only up to a constant of proportionality.). The cost
of the algorithm is quantified by the number of times $N$ the testing
circuit must be run to estimate $F(\mu,\tau)$ to precision $\epsilon$.
In the important regime in which $\mu$ is close to $\tau$, $N\approx4\epsilon^{-2}(1-F(\mu,\tau))dp_{\text{coll}}^{(\tau)}$
where $d$ is the dimension of the Hilbert space in which $\mu,\tau$
reside and $p_{\text{coll}}^{(\tau)}$ is the collision probability
of the target distribution. Note that $p_{\text{coll}}^{(\tau)}$
is a measure of the concentratedness of $\tau$ ($1/p_{\text{coll}}^{(\tau)}$
is the effective support of $\tau$). It follows that EVAQS is efficient
if $\tau$ is anticoncentrated. Optionally, an auxiliary quantum state
$\alpha$ may be used to importance sample $\mu$, further reducing
the number of repetitions needed. Any state $\alpha$ that is efficiently
preparable, well-characterized, and has support over $\tau$ may be
used. For this more general version of the algorithm I find that $N\approx4\epsilon^{-2}(1-F(\mu,\tau))(1+\chi^{2}(\tau,\alpha))$
where $\chi^{2}(\tau,\alpha)$ is the chi-square divergence between
the distributions induced by $\alpha$ and $\tau$. In this version,
the efficiency of EVAQS is limited only by how well $\alpha$ samples
$\tau$.

The state verification method proposed here constitutes a novel quantum
capability, as there is no method of verifying an anticoncentrated
\emph{classical} distribution with fewer than $O(d^{1/2})$ samples,
where $d$ is the effective support \citep{Valiant2014}\footnote{This conclusion is obtained by applying Theorem 1 of \citep{Valiant2014}
to the uniform distribution.}. And until recently, there was no known efficient method of verifying
arbitrary anticoncentrated quantum states. While this manuscript was
in preparation, Huang et al. introduced a type of randomized tomography
that enables low-rank projections of arbitrary quantum states (e.g.,
fidelity) to be estimated efficiently \citep{Huang2020}. In their
method the unknown state is measured in a set of random bases, each
obtained by applying a random global Clifford operation to the unknown
state. As they acknowledged, most global Clifford operations are non-trivial
to implement. Also, estimating a fidelity with their approach evidently
requires a substantial portion of the target state to be classically
computed. In contrast, the method proposed here involves only very
simple circuits and requires the calculation of only a small fraction
of the target state. On the other hand, the results of such calculations
must be available on demand for a feed-forward measurement. Regardless
of which method is ultimately found to be more practical, the approach
described here is novel and stands to be interesting in its own right.

While this work addresses a major problem (namely, sample complexity)
for verification of some large quantum states, it must be acknowledged
that some challenges remain. First of all, quantum states that have
an ``in between'' amount of concentration fall in the gap between
this work and prior work and remain difficult to verify. For example,
spin-glass thermal states can simultaneously have exponentially large
support (making them costly for standard methods) and be exponentially
sparse (making them difficult for the method described here). Secondly,
there is the very fundamental problem that verifying a quantum state
requires a specification of its readily measurable properties, whereas
the most useful quantum states for computational purposes are the
ones for which such a classical specification is likely to be difficult
to obtain. While it is perhaps too much to hope that all interesting
quantum states would be efficiently verifiable, this work shows that
some significant inroads may be made.

\section{Results}

\subsection{The EVAQS Algorithm}

\label{sec: EVAQS algorithm}Let $\ket{\mu}$ be an efficiently preparable
$n$-qubit state that is intended to approximate a (pure) target state
$\ket{\tau}=\sum_{x=1}^{d}\tau_{x}\ket x$ where $d=2^{n}$. Let us
call a projection of $\ket{\tau}$ onto a sparse subset of the computational
basis a \emph{snippet}. The EVAQS algorithm is motivated by two observations:
The first is that $\ket{\mu}$ and $\ket{\tau}$ are similar if and
only if all corresponding snippets of $\ket{\mu}$ and $\ket{\tau}$
are similar. The second is that, given the ability to compute selected
coefficients of $\ket{\tau}$, it is not difficult to prepare snippets
of $\ket{\tau}$ for direct comparison with corresponding snippets
of $\ket{\mu}$. This suggests the following strategy: (a) Project
out a random snippet of $\ket{\mu}$. (b) Construct the corresponding
snippet of $\ket{\tau}$ and test its fidelity with the snippet of
$\ket{\mu}$ using the quantum swap test \citep{Buhrman2006}. (c)
Repeat (a)-(b) many times and estimate the fidelity $F(\mu,\tau)\equiv\left|\braket{\mu}{\tau}\right|^{2}$
as a weighted average of the fidelities of the projected snippets.
In a nutshell, the idea is to verify a quantum state by verifying
its projections onto random two-dimensional subspaces.

Below I present two versions of the algorithm: A basic version that
samples $\mu$ uniformly and is efficient when $\tau$ is anticoncentrated,
and a more general version that uses an auxiliary quantum state $\alpha$
to importance sample $\mu$ when $\tau$ is not anticoncentrated.
In principle $\alpha$ can be any accurately characterized reference
state, but to be useful it should place substantial probability on
the support of $\tau$ while being significantly easier to prepare
than $\tau$ itself. The general version of the algorithm involves
a larger and more complex quantum circuit than the basic version,
but requires far fewer circuit repetitions when $\alpha$ is a substantially
better sampler of $\mu$ than the uniform distribution. Both versions
of the algorithm require the ability to compute, within the decoherence
time of the qubits, $\tau_{x}^{\prime}\propto\tau_{x}$ for any given
$x$; for the general version, the ability to compute $\alpha_{x}^{\prime}\propto\alpha_{x}$
is also required.

\begin{figure}
\begin{centering}
\includegraphics[scale=0.6]{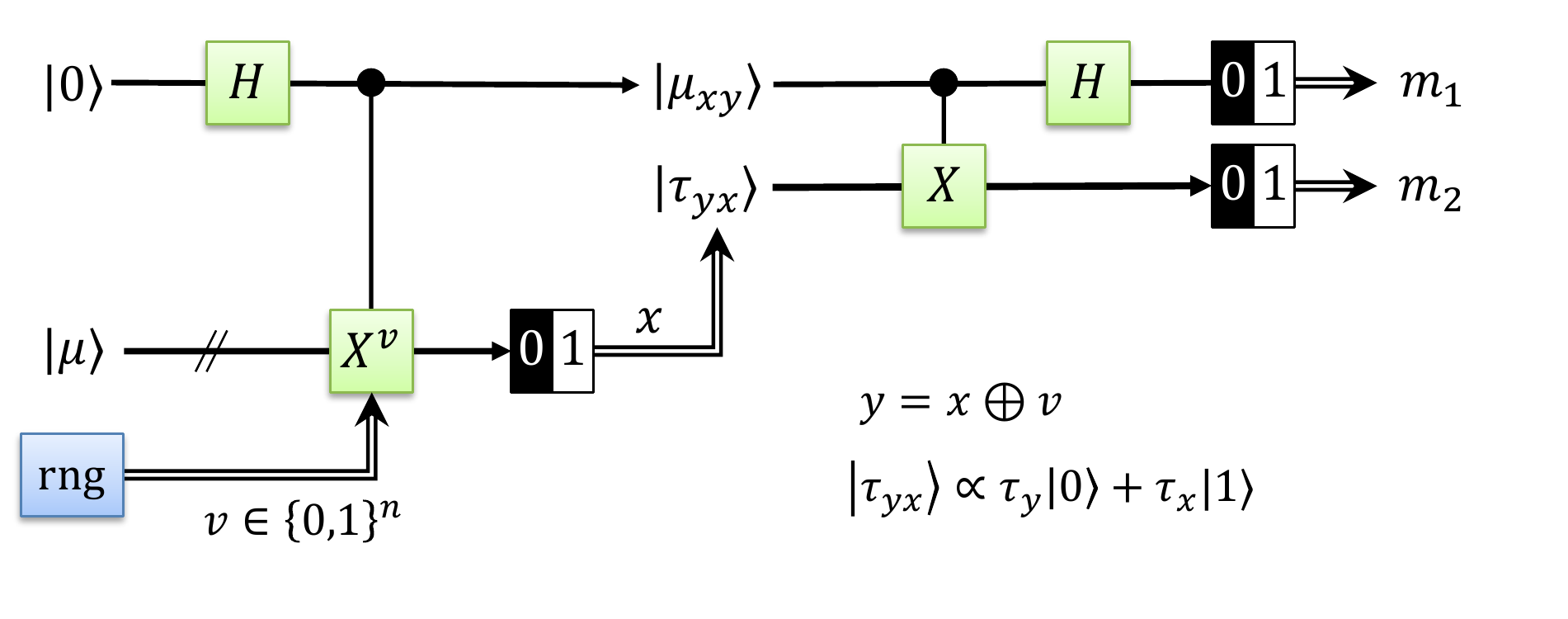}
\par\end{centering}
\vspace{1cm}
\begin{centering}
\includegraphics[scale=0.6]{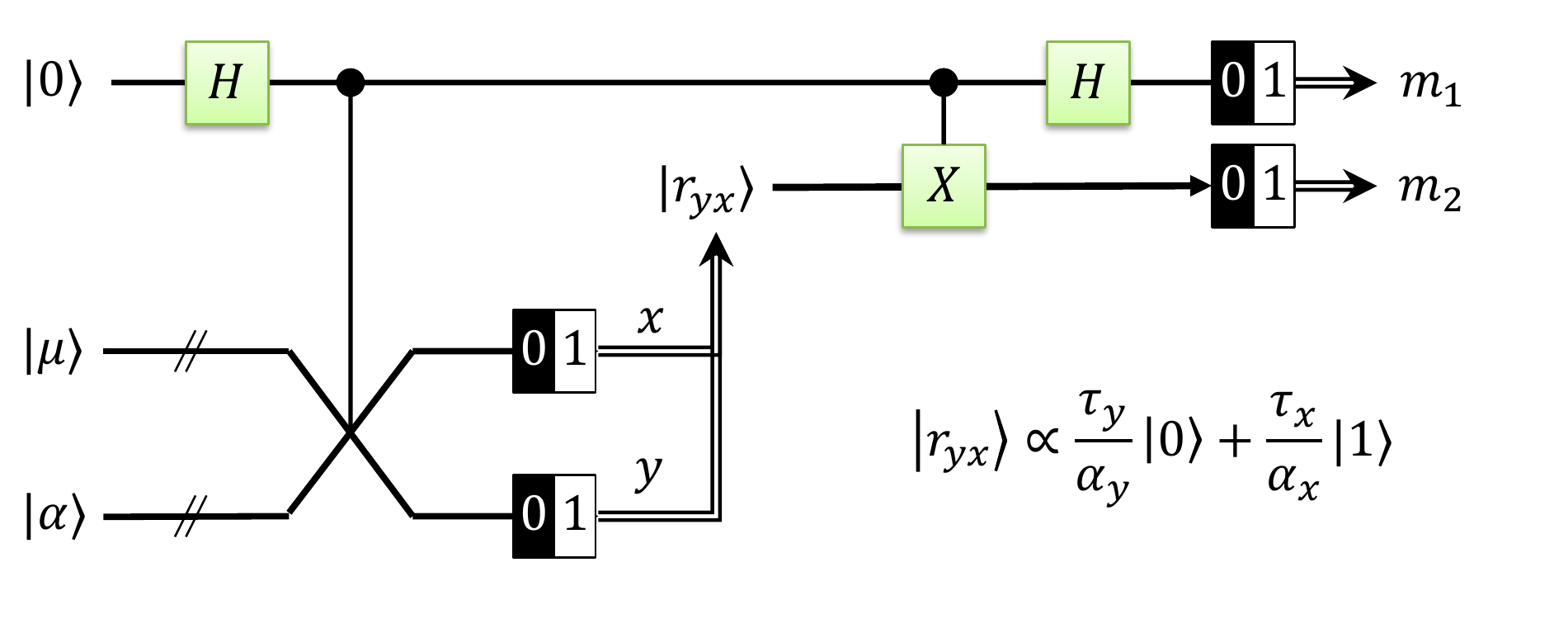}
\par\end{centering}
\caption{\label{fig: circuit}(top) A circuit to test an unknown quantum state
$\mu$ with respect to a classically-described target state $\tau$.
The circuit is repeated many times and the fidelity $F(\mu,\tau)$
between $\mu$ and $\tau$ is estimated in terms of weighted averages
of $b=(-1)^{m_{1}}m_{2}$ and $b^{2}$ (eq.~(\ref{eq: simple estimator})).
(bottom) An importance sampling circuit which reduces the number of
circuit repetitions needed when $\tau$ is not anticoncentrated. Here
$\alpha$ is any accurately-known state that samples $\mu$ and $\tau$
well. Notation: $H$ = Hadamard gate; $X$ = Pauli $X$ gate; 0/1
block = measurement in the computational basis; rng = random number
generator.}

\end{figure}

The basic version of the algorithm is implemented by a repeating a
short quantum-classical circuit involving a test register containing
the unknown state $\ket{\mu}\in\mathcal{H}_{2^{n}}$ and two ancilla
qubits (Fig.~\ref{fig: circuit} (top)). The steps of the circuit
is as follows:
\begin{enumerate}
\item Prepare the unknown state $\ket{\mu}$ and one of the ancillas in
the state $(\ket 0+\ket 1)/\sqrt{2}$.
\item Draw a uniformly distributed random number $v\in\{0,1\}^{n}$ and
perform controlled-NOT operations between the ancilla qubit and each
qubit $i$ in the test register for which $v_{i}=1$.
\item Measure the test register in the computational basis, obtaining a
value $x$.
\item Compute $\tau_{x}^{\prime}\propto\tau_{x}$ and $\tau_{y}^{\prime}\propto\tau_{y}$
where $y=x\oplus v$ ($\oplus$ denotes vector addition modulo 2).
\item Prepare the second ancilla qubit in the normalized state
\begin{align}
\ket{\tau_{yx}} & \propto\tau_{y}^{\prime}\ket 0+\tau_{x}^{\prime}\ket 1.
\end{align}
\item Measure the two ancilla qubits in the Bell basis and set $b=\pm1$
if $\ket{\Phi_{\pm}}\equiv(\ket 0\ket 0\pm\ket 1\ket 1)/\sqrt{2}$
is obtained, otherwise $b=0$. (Note that the Bell measurement can
be performed by a controlled-NOT between the ancillas followed by
a pair of single-qubit measurements, as shown in Fig. \ref{fig: circuit}a.)
\end{enumerate}
The general version of the algorithm uses an additional $n$-qubit
register containing an auxiliary state $\alpha$ Fig.~\ref{fig: circuit}
(bottom). In principle $\alpha$ can be any well-known quantum state,
though later it will be shown that the efficiency of the algorithm
directly depends on how well $\alpha$ samples both $\tau$ and the
error vector $\mu-\tau$. In this case the steps of the circuit are:
\begin{enumerate}
\item Prepare the test register, auxiliary register, and first ancilla qubit
in the state $\ket{\psi}=\frac{1}{\sqrt{2}}\ket{\mu}\ket{\alpha}(\ket 0+\ket 1)$.
\item Perform a controlled swap between the test and auxiliary registers,
using the first ancilla qubit as the control.
\item Measure the test and auxiliary registers in the computational basis,
obtaining values $x,y$ respectively.
\item Compute $\tau_{x}^{\prime}\propto\tau_{x}$, $\tau_{y}^{\prime}\propto\tau_{y}$,
$\alpha_{x}^{\prime}\propto\alpha_{x}$, and $\alpha_{y}^{\prime}\propto\alpha_{y}$.
\item Prepare a second ancilla qubit in the normalized state
\begin{align}
\ket{r_{yx}} & \propto\frac{\tau_{y}^{\prime}}{\alpha_{y}^{\prime}}\ket 0+\frac{\tau_{x}^{\prime}}{\alpha_{x}^{\prime}}\ket 1.
\end{align}
\item Same as for the basic version.
\end{enumerate}
For either version of the algorithm, steps 1-6 are repeated $N\gg1$
times. Let $x_{i},y_{i},b_{i}$ denote the values obtained in the
$i$th trial. Define the sample weight
\begin{align}
w_{x,y}^{\prime} & \equiv\left|\frac{\tau_{x}^{\prime}}{\alpha_{x}^{\prime}}\right|^{2}+\left|\frac{\tau_{y}^{\prime}}{\alpha_{y}^{\prime}}\right|^{2}
\end{align}
where $\alpha_{x}^{\prime}=2^{-n/2}$ for the basic version of the
algorithm. Then $F(\mu,\tau)$ is estimated by the quantity
\begin{align}
\tilde{F} & =\frac{\tilde{A}}{\tilde{B}}\label{eq: simple estimator}
\end{align}
where
\begin{align}
\tilde{A} & \equiv\frac{1}{N}\sum_{i=1}^{N}w_{x_{i},y_{i}}^{\prime}b_{i}\\
\tilde{B} & \equiv\frac{1}{N}\sum_{i=1}^{N}w_{x_{i},y_{i}}^{\prime}b_{i}^{2}.
\end{align}
The simple estimator (\ref{eq: simple estimator}) has a bias of order
$N^{-1}$. In practice this is usually negligible, but for completeness
an estimator that is unbiased to order $N^{-1}$ is derived in Appendix
\ref{sec: bias correction}.

I note that the circuit for the basic algorithm is quite simple, consisting
of (on average) just $n/2+1$ controlled-NOT gates and a few single-qubit
gates and measurements. The general version replaces the (on average)
$n/2$ controlled-NOT gates with $n$ controlled-SWAP (Fredkin) gates
between $n$ pairs of qubits. Each controlled-SWAP gate requires a
handful of standard 1- and 2-qubit gates \citep{Smolin1996,Hung2006}.
Thus the general version uses twice as many qubits and roughly 10
times as many gates as the basic version. However, if $\alpha$ is
substantially better at sampling $\tau$ than the uniform distribution,
the general version will require substantially fewer circuit repetitions
to obtain a precise estimate.

\subsection{Performance}

\subsubsection{Cost to Achieve a Given Precision}

The cost of EVAQS is driven by the number of samples needed to obtain
an estimate with sufficiently small variance. In a typical application
the expectation is that $\mu$ is not a bad approximation of $\tau$;
thus the regime of greatest interest is that of small infidelity $I\equiv1-F$.
To lowest order in $I$ and statistical fluctuations, the variance
of $\tilde{F}$ can be written as
\begin{align}
\Var\tilde{F} & \approx\frac{2}{N}\sum_{x}\frac{I\left|\tau_{x}\right|^{4}+\left|\tau_{x}\right|^{2}\left|\epsilon_{x}\right|^{2}}{\left|\alpha_{x}\right|^{2}}.\label{eq: variance expression (epsilon)}
\end{align}
where $\epsilon\equiv e^{-\imag\phi}\mu-\tau$ is the error of $\mu$
upon correcting its global phase $\phi$, defined via $\braket{\tau}{\mu}=\sqrt{F}e^{\imag\phi}$
. As shown in the Appendix, $\left|\epsilon_{x}\right|^{2}$ is (to
lowest order) also proprtional to $I$, so that the entire expression
is proportional to $I$. Note that $\Var\tilde{F}$ does not depend
on the phases of the components of $\alpha$.

The worst case is that the error $\epsilon$ resides entirely on the
component $x$ for which $\left|\tau_{x}/\alpha_{x}\right|$ is largest,
i.e. the basis state which is most badly undersampled. In non-adversarial
scenarios the error may be expected to be distributed across the support
of $\tau$. Simulations indicate that for plausible error models $\sum_{x}\left|\tau_{x}\right|^{2}\left|\epsilon_{x}\right|^{2}/\left|\alpha_{x}\right|^{2}$
is typically comparable in magnitude to $I\sum_{x}\left|\tau_{x}\right|^{4}/\left|\alpha_{x}\right|^{2}$.
This leads to the scaling heuristic
\begin{align}
N & \sim\frac{4I}{\epsilon^{2}}\left(1+\chi^{2}(\left|\tau\right|^{2},\left|\alpha\right|^{2})\right)\label{eq: cost estimate, chi-2}
\end{align}
where $\epsilon^{2}$ is the desired variance and $\chi^{2}(p,q)\equiv\sum_{x}\left|p_{x}-q_{x}\right|^{2}/q_{x}$
is the chi-square divergence of $p$ with respect to $q$. $\chi^{2}$
is a standard statistical measure that quantifies the ``distance''
between distributions $p$ and $q$, heavily weighting differences
in which $q$ undersamples $p$.\footnote{$\chi^{2}$ can be understood as the lowest-order contribution to
the Kullback-Liebler divergence, a similar well-known measure with
useful information-theoretic properties.} Note that $N$ has no intrinsic dependence on the dimension of $\tau$.

For the basic version of the algorithm (or when $\alpha$ is the uniform
superposition), eq.~(\ref{eq: cost estimate, chi-2}) can be simplified
even further to
\begin{align}
N & \sim\frac{4I}{\epsilon^{2}}dp_{\text{coll}}^{(\tau)}\label{eq: cost estimate, uniform}
\end{align}
where $p_{\text{coll}}^{(\tau)}=\sum_{x}\left|\tau_{x}\right|^{4}$
is the collision probability of the classical distribution induced
by $\tau$. $1/p_{\text{coll}}^{(\tau)}$ may be interpreted as the
effective support size $d_{\text{eff}}$ of $\tau$. It follows that
the basic algorithm is efficient so long as $d_{\text{eff}}/d$ is
at least $1/\text{poly}(n)$, that is, so long as the effective support
of $\tau$ is not too small. This make the proposed method complementary
to existing methods which are efficient for concentrated distributions.
For a slightly different perspective, the factor $dp_{\text{coll}}^{(\tau)}$
can be written as $2^{n-H_{2}^{(\tau)}}$ where $H_{2}^{(\tau)}$
is the R�nyi 2-entropy of the classical distribution induced by $\tau$.
Thus the basic method is efficient when the entropy is large, namely,
when $n-H_{2}^{(\tau)}$ grows at most logarithmically in $n$.

Using standard calculus one can show that the optimal sampling distribution
satisfies
\begin{align}
\left|\alpha_{x}\right|^{2} & \propto\sqrt{\left|\tau_{x}\right|^{2}\left(\left|\tau_{x}\right|^{2}+\left|\sigma_{x}\right|^{2}\right)}
\end{align}
where $\sigma$ is the normalized projection of $\epsilon$ onto the
space orthogonal to $\tau$. One might have guessed that it would
be sufficient for $\alpha$ to sample either $\mu$ or $\tau$ well.
But that is incorrect: If $\alpha$ doesn't frequently sample the
components of $\mu$ with the largest errors (whether or not the components
themselves are large), then the (in)fidelity will not be estimated
with high precision. 

\subsubsection{Robustness to Error in the Auxiliary State}

\label{sec: robustness}A key assumption in the general version of
EVAQS is that the auxiliary state $\alpha$ is well-characterized.
Fortunately, EVAQS is robust with respect to this assumption in the
sense that small error in the knowledge of $\alpha$leads to a correspondingly
small error in the estimate of $F(\mu,\tau)$.

Suppose $\alpha$ is mistakenly characterized as $\tilde{\alpha}$.
Then as shown in Section \ref{sec: robustness derivation}, instead
of estimating $F(\mu,\tau)$ the procedure estimates $F(\mu,\tilde{\tau}$)
where $\tilde{\tau}$ is a perturbed version of $\tau$, namely
\begin{align}
\tilde{\tau}_{x} & \equiv\frac{\tau_{x}\alpha_{x}/\tilde{\alpha}_{x}}{\sqrt{\sum_{y}\left|\tau_{y}\alpha_{y}/\tilde{\alpha}_{y}\right|^{2}}}.
\end{align}
Intuitively, if $\tilde{\alpha}$ is close to $\alpha$ then $\tilde{\tau}$
will be close to $\tau$ and $F(\mu,\tilde{\tau})$ will be close
to $F(\mu,\tau)$. In Section \ref{sec: robustness derivation} and
Appendix \ref{sec: robustness bound derivation} it is shown that
\begin{align}
F(\mu,\tau) & \ge F(\mu,\tilde{\tau})-\frac{2\delta_{\text{rms}}}{1-\delta_{\text{rms}}}\label{eq: robustness bound}
\end{align}
where $\delta_{\text{rms}}$ is the average relative error in $\tilde{\alpha}$.
Thus a small error in the characterization of $\alpha$ yields a correspondingly
small error in the estimate of $F(\mu,\tau)$.

\subsection{Simulations}

\label{sec: simulations}In this section I present the results of
several simulation studies demonstrating the validity and efficacy
of the proposed method. The first study demonstrates scalable verification
of so-called instantaneous quantum polynomial (IQP) circuits \citep{Shepherd2009}.
The next two studies demonstrate sample-efficient verification of
random quantum circuits, including the kind of circuits used to demonstrate
quantum supremacy \citep{Arute2019}. In each of these studies the
basic version of the method (without an auxiliary state $\ket{\alpha}$)
was employed. 

\subsubsection{Verification of IQP Circuits}

IQP circuits are currently of interest as a family of relatively simple
quantum circuits whose output distributions are hard to simulate classically
\citep{Bremner2011,Bremner2016}. IQP circuits are well-suited for
demonstrating EVAQS as their outputs are typically anticoncentrated
in the computational basis. Even better, if the output an IQP circuit
is transformed into the Hadamard basis, the distribution is not only
perfectly uniform (yielding the lowest possible sample complexity),
but also easy to calculate on a classical computer. This makes EVAQS
a fully scalable way to verify IQP circuits.

An $n$-qubit IQP circuit of depth $m$ can be defined as a set of
$m$ multiqubit $X$ rotations acting on the $\ket 0^{n}$ state.
A multiqubit NOT operation may be written as $X^{a}\equiv X_{1}^{a_{1}}\otimes\cdots\otimes X_{n}^{a_{n}}$
for $a\in\{0,1\}^{n}$. In terms of such operators, an IQP circuit
has the form 
\begin{align}
\ket{\tau} & =\exp\left(\imag\sum_{i=1}^{m}\theta_{i}X^{A_{i}}\right)\ket 0^{n}.
\end{align}
for some set of vectors $A_{1},\ldots,A_{m}\in\{0,1\}^{n}$. The amplitudes
of the output state can be written concisely as
\begin{align}
\tau_{x} & =\braket x{\tau}\\
 & =\sum_{v:~Av=x}\beta_{1}(v_{1})\cdots\beta_{n}(v_{n})
\end{align}
where $A=[A_{1},\ldots,A_{m}]$ and
\begin{align}
\beta_{i}(b) & =\begin{cases}
\cos\theta_{i} & b=0\\
\imag\sin\theta_{i} & b=1
\end{cases}.
\end{align}
The number of terms in $\sum_{v:~Av=x}$ is $2^{m-r}$ where $r=\rank(A)$.
Since $r\le n$, the number of terms contributing to $\tau_{x}$ is
exponential in the circuit depth $m$ once it exceeds $n$.

\paragraph{In the Hadamard Basis}

IQP states are substantially easier to analyze in the Hadamard basis.
Let $\ket{\xi}=H^{\otimes n}\ket{\tau}$. Note that it is experimentally
easy to obtain $\ket{\xi}$ from $\ket{\tau}$. Since $HX=ZH$, we
have
\begin{align}
\ket{\xi} & =\exp\left(\imag\sum_{i=1}^{m}\theta_{i}Z^{A_{i}}\right)\ket +^{n}
\end{align}
where $\ket +\equiv(\ket 0+\ket 1)/\sqrt{2}=H\ket 0$. In this basis,
the amplitude
\begin{align}
\braket x{\xi} & =\frac{1}{\sqrt{2^{n}}}\exp\left(\imag\sum_{i=1}^{m}\theta_{i}(-1)^{x\cdot A_{i}}\right)
\end{align}
is trivial to compute classically. Furthermore, the induced probability
distribution is uniform, which is the best case for EVAQS.

As a first demonstration, I simulated the verification of random IQP
circuits in the Hadamard basis. The test circuits were comprised of
$n=4,6,8,\ldots,20$ qubits and $m=3n$ rotations; the depth $m=3n$
was chosen to ensure that the complexity of the output state is exponential
in $n$. For each rotation, a random subset of qubits was chosen from
$n$ Bernoulli trials such that on average 2 qubits were involved
in each rotation. Each rotation angle was chosen uniformly from $[0,2\pi]$.
To simulate circuit error, each angle $\theta_{i}$ was perturbed
by a small random amount $\delta_{i}$. The angle errors were globally
scaled so that the resulting state $\ket{\mu}$ had a prescribed infidelity
$I$ with respect to the ideal state. For each $n$ and each $I\in\{0.01,0.03,0.1,0.3\}$,
400 random noisy circuits were realized. For each circuit, the verification
procedure was performed with $10^{4}$ simulated measurements.

The left plot in Fig.~\ref{fig: IQP simulation, hadamard} shows
the estimated infidelities obtained from these simulated experiments.
(In all the figures, a solid line shows the median value among all
realizations of a given experiment and a surrounding shaded band shows
the 10th-90th percentiles). As expected, the proposed method is able
to accurately estimate the infidelity of the prepared state independent
of number of qubits and over a wide range of infidelities. With a
fixed number of measurements, the states with smaller infidelity are
estimated with larger relative error; however, the absolute error
is actually smaller, in accordance with eq. (\ref{eq: cost estimate, uniform}).
The right panel of Fig.~\ref{fig: IQP simulation, hadamard} shows
the sample cost of the procedure, given by eq. (\ref{eq: true variance})
and normalized by the desired precision $\epsilon^{2}$, as a function
of the number of qubits. Notably, the cost is independent of the number
of qubits, depending only on the fidelity of the test state. Again,
this is expected from eq. (\ref{eq: cost estimate, uniform}) given
that the output distribution is uniform.

\begin{figure}
\begin{centering}
\includegraphics{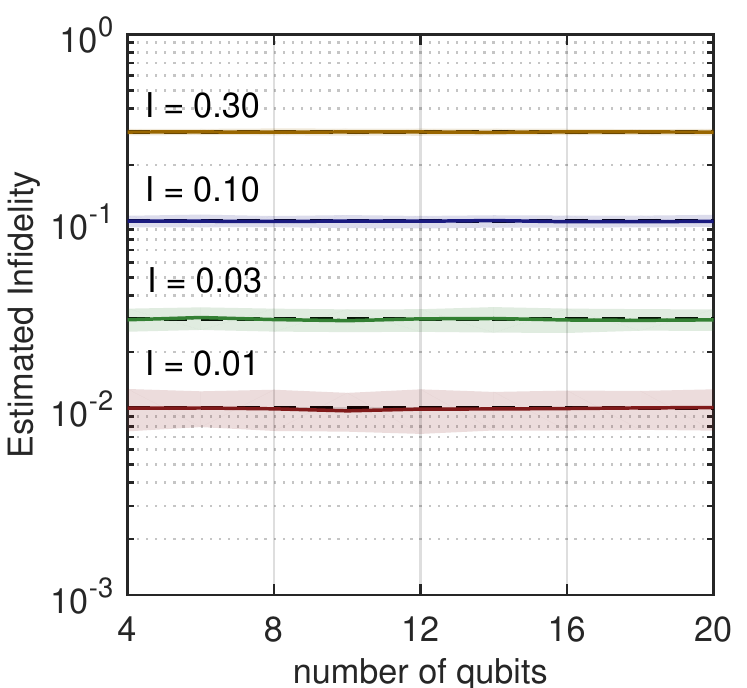}\includegraphics{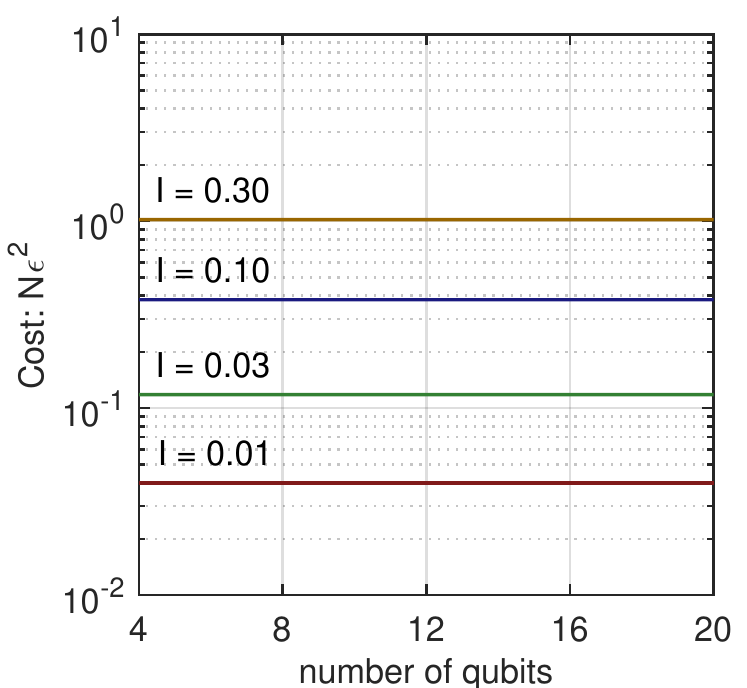}
\par\end{centering}
\caption{\label{fig: IQP simulation, hadamard}Simulated verification of IQP
circuits with rotation angle errors, using measurements in the Hadamard
basis. (left) Estimated infidelity. (right) Precision-normalized cost. }

\end{figure}

\paragraph{In the Computational Basis}

EVAQS is also effective when measurements are performed in the computational
basis. In this basis, the output state of a typical IQP circuit is
anticoncentrated in the sense that significant fraction of the basis
states have probabilities of order $2^{-n}$ or larger \citep{Webb2020}.
Fig. \ref{fig: IQP simulation, computational},left shows the estimated
infidelities for the same set of circuits as described in the previous
subsection, but this time using simulated measurements in the computational
basis. As before, EVAQS was able to accurately estimate the circuit
fidelities. This time, however, the cost tends to increase slowly
with the number of qubits. I note that while the median cost does
appear to grow exponentially, in going from 4 to 20 qubits it increases
only by a factor of about 2.5, whereas the size of the state being
verified increases by a factor of $2^{20}/2^{4}=65\thinspace536$.
Furthermore, there is noticeable variation in cost for circuits of
the same size (Fig.~\ref{fig: IQP simulation, computational}, middle).
This is because different random circuits of the same size were anticoncentrated
to different degrees. The strong link between cost and the degree
of anticoncentration in the target distribution is shown in Fig.~\ref{fig: IQP simulation, computational},right.
Indeed, the degree of concentration (as measured by the inverse collision
probability) is a better predictor of cost than the number of qubits.
Also noteworthy is the fact that the estimated costs, given by eq.
(\ref{eq: cost estimate, uniform}) and shown as dashed lines, are
within a small factor of the true costs.

\begin{figure}
\begin{centering}
\includegraphics{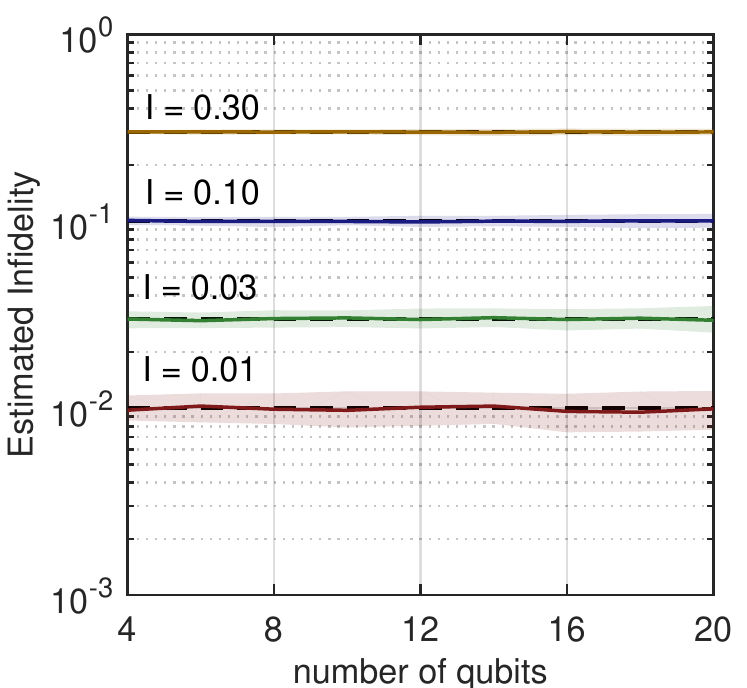}\includegraphics{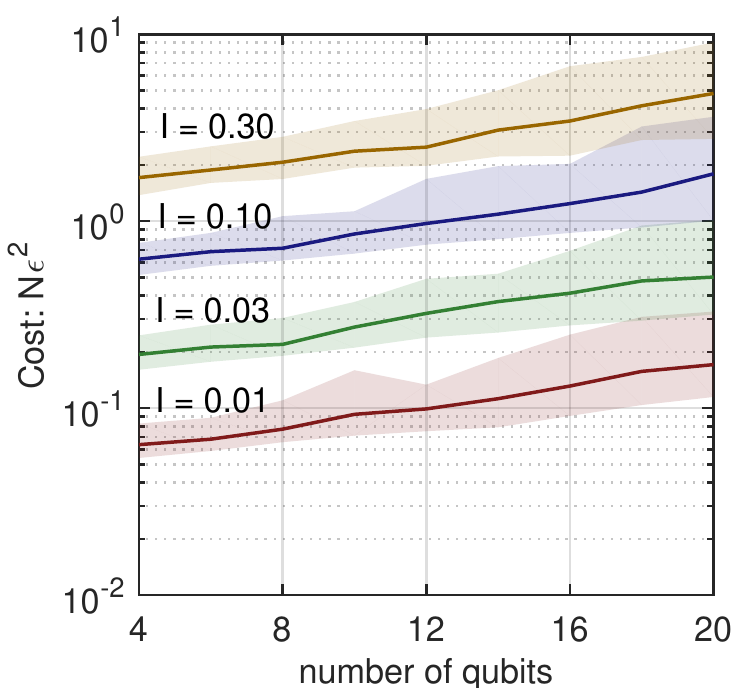}
\par\end{centering}
\begin{centering}
\includegraphics{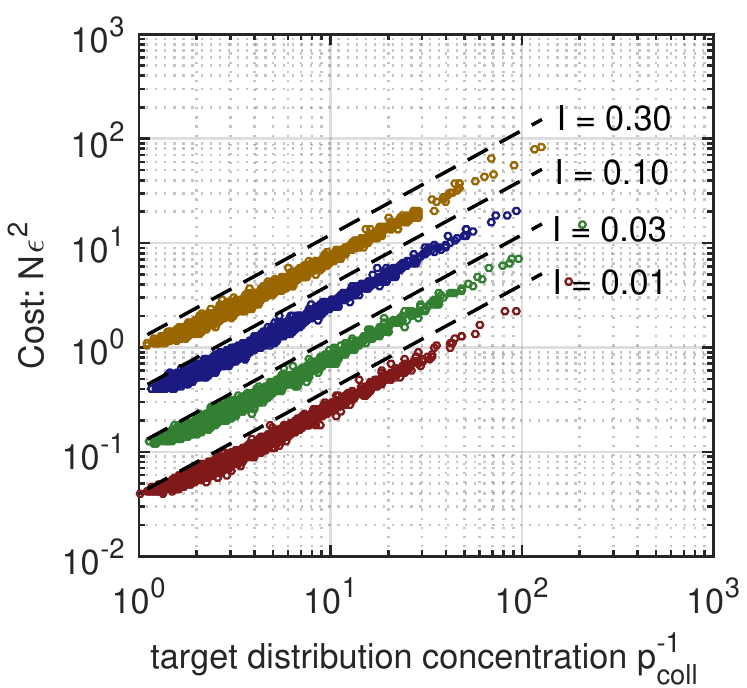}
\par\end{centering}
\caption{\label{fig: IQP simulation, computational}Simulated verification
of IQP circuits with rotation angle errors, using measurements in
the computational basis. (left) Estimated infidelity. (middle) Precision-normalized
cost. (right) Cost as a function of the concentratedness of the output
state. The dashed lines are the estimated cost, eq. (\ref{eq: cost estimate, chi-2}).}
\end{figure}

\subsubsection{Verification of Random Circuits}

The second study involves verification of random quantum circuits,
that is, sequences of random 2-qubit unitaries on randomly selected
pairs of qubits. Each random unitary was obtained by generating a
random complex matrix with normally-distributed elements, then performing
Gram-Schmidt orthogonalization on the columns of the matrix. In this
study, error was modelled as a perturbation of the output state rather
than perturbation of the individual gates. The output state was written
as
\begin{align}
\ket{\mu} & =\sqrt{1-\eta}\ket{\tau}+\sqrt{\eta}\ket{\epsilon}
\end{align}
where $\ket{\epsilon}$ consisted of both multiplicative and additive
noise,
\begin{align}
\epsilon_{x} & \sim\xi_{x}^{\prime}\tau_{x}+\lambda\xi_{x}^{\prime\prime}
\end{align}
where $\xi_{x}^{\prime},\xi_{x}^{\prime\prime}$ are independent complex
Gaussian random variables.The constant $\lambda$ was chosen so that
the standard deviations of the multiplicative error and additive error
are equal when $\left|\tau_{x}\right|$ equals its mean value. The
constant $\eta$ was then chosen to yield a particular infidelity
$I$. As before, circuits were comprised of $n=2,4,6,\ldots,20$ qubits
and $m=3n$ gates. For each $n$ and $I\in\{0.01,0.03,0.1,0.3\}$,
300 random circuits were realized. For each circuit, the verification
procedure was performed with $10^{4}$ simulated measurements.

\begin{figure}
\begin{centering}
\includegraphics{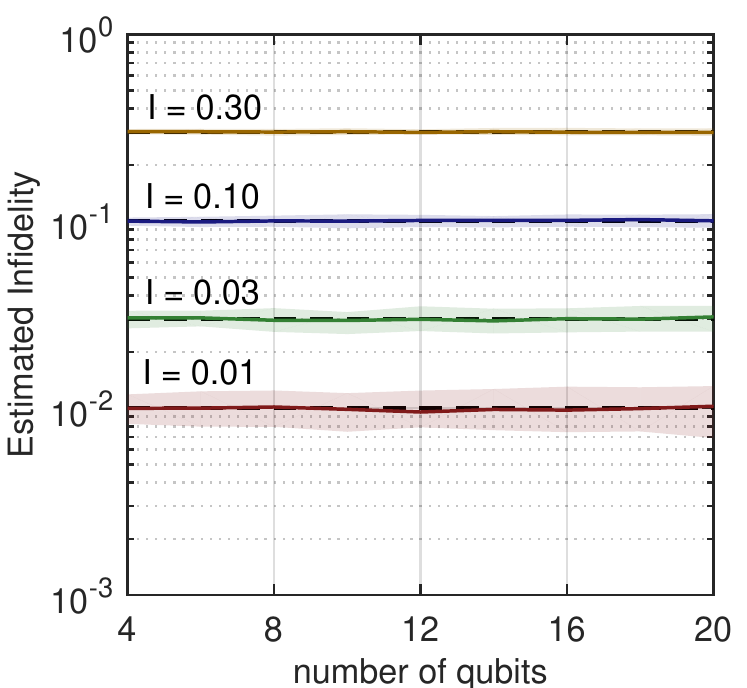}\includegraphics{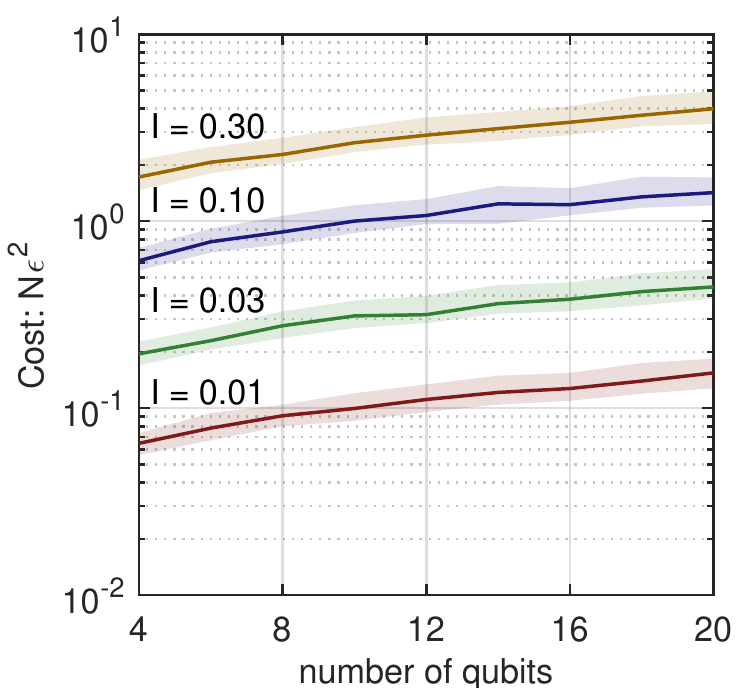}
\par\end{centering}
\begin{centering}
\includegraphics{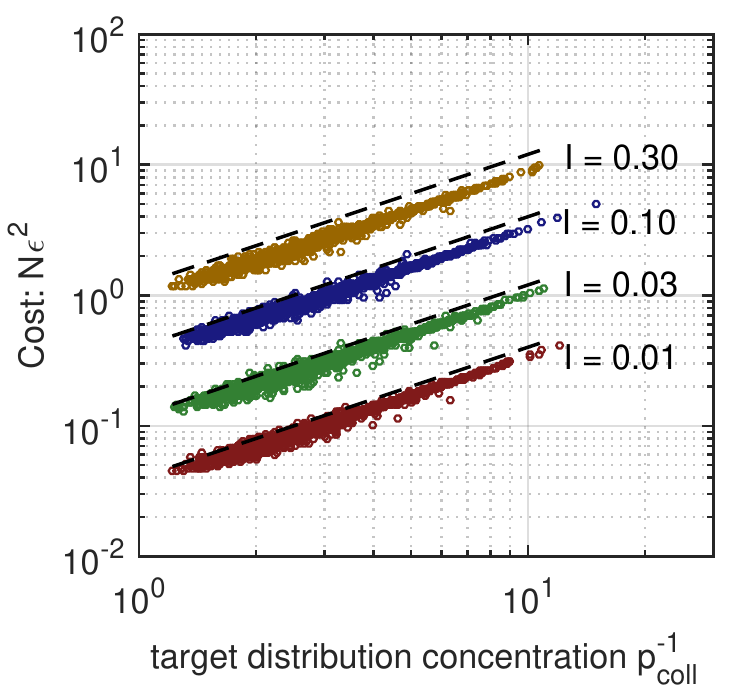}
\par\end{centering}
\caption{\label{fig: random circuit simulation}Simulated verification of quantum
circuits consisting of random 2-qubit unitaries and random error on
the output state. (left) Estimated infidelity. (middle) Precision-normalized
cost. (right) Cost as a function of the concentratedness of the output
state. The dashed lines are the estimated cost, eq. (\ref{eq: cost estimate, chi-2}).}

\end{figure}

The results are shown in Fig. \ref{fig: random circuit simulation}.
Again, EVAQS is able to estimate the output state fidelity accurately
with a number of measurements that grows very slowly with the number
of qubits. And as with IQP circuits, the cost is well-predicted by
the concentration of the target state and the infidelity of the unknown
state.

\subsubsection{Verification of Supremacy Circuits}

Another important class of circuits is that recently used to demonstrate
the ``quantum supremacy'' of a quantum processor \citep{Arute2019}.
Such circuits consist of alternating rounds of single-qubit rotations
drawn from a small discrete set and entangling operations on adjacent
pairs of qubits in a particular pattern. Like IQP circuits, such circuits
are hard to classically simulate \citep{Aaronson2016,Aaronson2020a}
in spite of their locality constraints. But unlike IQP circuits, they
are universal for quantum computing.

The circuits simulated in this study consisted of $n\in\{4,9,12,16,20\}$
qubits arranged in a planar square lattice with (nearly) equal sides
and 16 cycles of alternating single-qubit operations and entangling
operations, as described in \citep{Arute2019}. Single-qubit error
was modelled as a post-operation unitary of the form $\exp\left(\imag(\epsilon_{x}X+\epsilon_{y}Y+\epsilon_{z}Z\right)$
where $\epsilon_{x},\epsilon_{y},\epsilon_{z}$ are independent normally-distributed
variables of zero mean and small variance. Error on the two-qubit
operations was modelled as a small random perturbation of the angles
$\theta,\phi$ parameterizing the entangling gate \citep{Arute2019}.
The amount of error was chosen to yield a process infidelity \citep{Gilchrist2005}
on the order of 0.02\% per single-qubit operation and 0.2\% per two-qubit
operation. For each circuit size, 100 random noisy circuits were generated;
for each circuit, $10^{4}$measurements were simulated.

\begin{figure}
\begin{centering}
\includegraphics{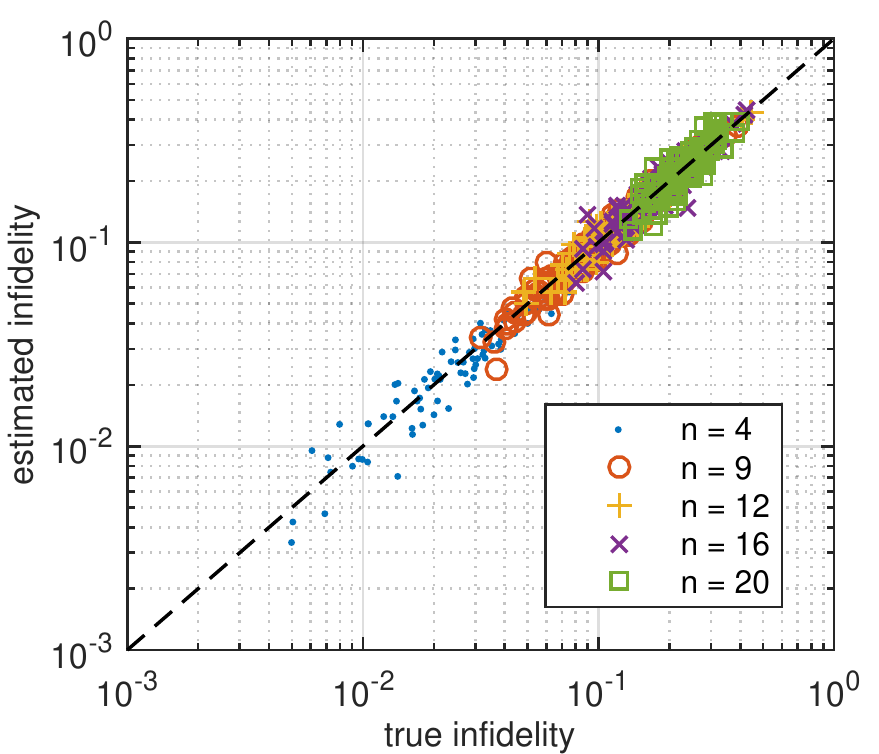}
\par\end{centering}
\caption{\label{fig: supremacy simulations}(color online) Simulated verification
of the types of circuits used in recent quantum supremacy demonstrations.
$n$ is the number qubits.}

\end{figure}

Fig.~\ref{fig: supremacy simulations} plots the estimated fidelity
vs. the true infidelity of all the circuits simulated. For comparison,
the dashed line shows the true infidelity. The smallest circuits ($n=4$)
had output infidelities on the order of $10^{-3}$, while the largest
($n=20$) had infidelities around 0.3. Over this range, the infidelity
was estimated to within 20\% or better. What is not evident from the
plot is that the variance of the estimator varied by an order of magnitude
for different random circuits of the same size, due to the different
amounts of entropy in their output distributions. Consequently, the
expected error for some of the circuits is actually considerably smaller
than 20\%. For the circuits that were verified less accurately, the
expected error could be reduced further by increasing the number of
measurements. Interestingly, the relative error does not vary much
over the wide range of circuit sizes and circuit infidelities in these
simulations. According to eq.~(\ref{eq: cost estimate, uniform}),
one expects the relative error to scale as $\sqrt{\Var\tilde{F}}/I\propto I^{-1/2}$,
that is, to increase as infidelity decreases. Additional simulations
confirmed that this scaling does occur with smaller gate errors.

\section{Methods}

\subsection{Proof of Correctness}

\label{sec: algorithm derivation}To see that the procedures described
in Section \ref{sec: EVAQS algorithm} indeed yield an estimate of
$F(\mu,\tau)$, let us consider the general case first; the correctness
of the simpler version will then be established as a special case.

\subsubsection{Proof of General Version}

In each iteration of the general algorithm, one first prepares the
test register, auxiliary register, and the first ancilla qubit in
the state
\begin{align}
\ket{\psi} & =\frac{1}{\sqrt{2}}\ket{\mu}\ket{\alpha}(\ket 0+\ket 1)\\
 & =\frac{1}{\sqrt{2}}\sum_{x,y=1}^{d}\mu_{x}\alpha_{y}\ket x\ket y(\ket 0+\ket 1).
\end{align}
In the second step, the ancilla controls a swap between the test and
auxiliary registers. This yields the state
\begin{align}
\ket{\psi^{\prime}} & =\frac{1}{\sqrt{2}}\sum_{x,y=1}^{d}\ket x\ket y(\mu_{x}\alpha_{y}\ket 0+\mu_{y}\alpha_{x}\ket 1).\label{eq: state after CSWAP}
\end{align}
In the third step the test and auxiliary registers are measured, yielding
values $x,y$. This projects ancilla 1 onto the unnormalized state
\begin{align}
\ket{\psi_{xy}} & =\frac{1}{\sqrt{2}}(\mu_{x}\alpha_{y}\ket 0+\mu_{y}\alpha_{x}\ket 1).\label{eq: measured snippet}
\end{align}
In the fourth step, the observed values $x,y$ are used to compute
$\tau_{x}^{\prime},\tau_{y}^{\prime},\alpha_{x}^{\prime},\alpha_{y}^{\prime}$
and prepare ancilla qubit 2 in the state
\begin{align}
\ket{r_{yx}} & \equiv\frac{1}{\sqrt{w_{yx}^{\prime}}}\left(\frac{\tau_{y}^{\prime}}{\alpha_{y}^{\prime}}\ket 0+\frac{\tau_{x}^{\prime}}{\alpha_{x}^{\prime}}\ket 1\right).
\end{align}
Let $\lambda=\left|\tau_{x}^{\prime}/\tau{}_{x}\right|^{2}/\left|\alpha_{x}^{\prime}/\alpha_{x}\right|^{2}$,
which is independent of $x$. Then $\ket{r_{yx}}$ can be written
as
\begin{align}
\ket{r_{yx}} & =\frac{1}{\sqrt{w_{yx}}}\left(\frac{\tau_{y}}{\alpha_{y}}\ket 0+\frac{\tau_{x}}{\alpha_{x}}\ket 1\right)
\end{align}
where $w_{xy}=w_{xy}^{\prime}/\lambda$. The joint state of the two
ancillas is
\begin{align}
\ket{\psi_{xy}^{\prime}} & =\frac{1}{\sqrt{2w_{xy}}}\left(\mu_{x}\alpha_{y}\ket 0+\mu_{y}\alpha_{x}\ket 1\right)\left(\frac{\tau_{y}}{\alpha_{y}}\ket 0+\frac{\tau_{x}}{\alpha_{x}}\ket 1\right)\label{eq: measured snippet with ancilla}
\end{align}
Finally, the ancillas are measured in the Bell basis. Since
\begin{align}
\braket{\Phi_{\pm}}{\psi_{xy}^{\prime}} & =\frac{1}{\sqrt{2w_{xy}}}\left(\mu_{x}\tau_{y}\pm\mu_{y}\tau_{x}\right)\label{eq: Bell state projection}
\end{align}
the probability of outcome $b=\pm1$ is
\begin{align}
p_{x,y,\pm1} & =\frac{1}{4w_{xy}}\left(\left|\mu_{x}\right|^{2}\left|\tau_{y}\right|^{2}+\left|\mu_{y}\right|^{2}\left|\tau_{x}\right|^{2}\pm\mu_{x}\tau_{y}\mu_{y}^{*}\tau_{x}^{*}\pm\mu_{y}\tau_{x}\mu_{x}^{*}\tau_{y}^{*}\right).
\end{align}
Using $\left|\sum_{x}\mu_{x}^{*}\tau_{x}\right|^{2}=F(\mu,\tau)$
we obtain
\begin{align*}
\sum_{x,y}w_{xy}p_{xy\pm} & =\frac{1}{2}\left(1\pm F(\mu,\tau)\right).
\end{align*}
Now, $p_{xy+}+p_{xy-}=\sum_{b\in\{-1,0,1\}}p_{xyb}b^{2}$ and $p_{xy+}-p_{xy-}=\sum_{b\in\{-1,0,1\}}p_{xyb}b$.
Thus
\begin{align}
\left<w_{xy}^{\prime}b\right> & =\lambda\sum_{x,y,b}w_{xy}p_{xyb}b\\
 & =\lambda\sum_{x,y}w_{xy}(p_{xy+}-p_{xy-})\\
 & =\lambda F(\mu,\tau)
\end{align}
and
\begin{align}
\left<w_{xy}^{\prime}b^{2}\right> & =\lambda\sum_{x,y,b}w_{xy}p_{xyb}b^{2}\\
 & =\lambda\sum_{x,y}w_{xy}(p_{xy+}+p_{xy-})\\
 & =\lambda.
\end{align}
Thus
\begin{align}
F(\mu,\tau) & =\frac{\left<w_{xy}^{\prime}b\right>}{\left<w_{xy}^{\prime}b^{2}\right>}.
\end{align}

To obtain an experimental estimate of $F(\mu,\tau)$, the steps above
are repeated $N\gg1$ times. Let $x_{i},y_{i},b_{i}$ denote the values
of $x,y,b$ obtained in the $i$th trial. Let $A_{i}=w_{x_{i}y_{i}}^{\prime}b_{i}$
and $B_{i}=w_{x_{i}y_{i}}^{\prime}b_{i}^{2}$. The experimental quantities
\begin{align}
\tilde{A} & \equiv\frac{1}{N}\sum_{i=1}^{N}A_{i}\label{eq: A estimator}\\
\tilde{B} & \equiv\frac{1}{N}\sum_{i=1}^{N}B_{i}\label{eq: B estimator}
\end{align}
are unbiased estimators of $A=\lambda F$ and $B=\lambda$ respectively.
As a first approximation $F$ may be estimated by the ratio $\tilde{A}/\tilde{B}$.
It is a basic result of statistical analysis that such an estimator
has a bias of order $N^{-1}$. The estimator which corrects for this
bias is derived in Appendix \ref{sec: bias correction}.

\subsubsection{Proof of the Basic Version}

\label{sec: correctness - no auxiliary}To establish the correctness
of the basic version of the algorithm, I show that it is equivalent
to performing the general algorithm with $\alpha$ as the uniform
superposition state.

In the basic version, one starts with just the unknown state $\ket{\mu}$
and an ancilla qubit in the state $(\ket 0+\ket 1)/\sqrt{2}$. One
picks a uniform random vector $v\in\{0,1\}^{n}$ and performs
\begin{align}
\ket x\ket 0 & \rightarrow\ket x\ket 0\\
\ket x\ket 1 & \rightarrow\ket{x\oplus v}\ket 1
\end{align}
yielding the state
\begin{align}
\ket{\psi_{xy,x\oplus v}^{\prime}} & =\frac{1}{\sqrt{2}}\sum_{x}\mu_{x}\left(\ket x\ket 0+\ket{x\oplus v}\ket 1\right)\\
 & =\frac{1}{\sqrt{2}}\sum_{x}\ket x\left(\mu_{x}\ket 0+\mu_{x\oplus v}\ket 1\right).
\end{align}
Measuring $x$ yields the ancilla state
\begin{equation}
\mu_{x}\ket 0+\mu_{x\oplus v}\ket 1\label{eq: ancilla state (RNG)}
\end{equation}
with net probability
\begin{align}
p_{x,v} & =\frac{1}{2^{n+1}}\left(\left|\mu_{x}\right|^{2}+\left|\mu_{x\oplus v}\right|^{2}\right).\label{eq: p_xy (RNG)}
\end{align}

Now, consider the general algorithm with $\ket{\alpha}=2^{-n/2}\sum_{y}\ket y$.
From eq.\,(\ref{eq: state after CSWAP}), the state immediately following
the controlled swap between test and auxiliary registers is
\begin{align}
\ket{\psi^{\prime}} & =\frac{1}{\sqrt{2^{n+1}}}\sum_{x,y}\ket x\ket y(\mu_{x}\ket 0+\mu_{y}\ket 1).
\end{align}
Measurement of $x,y$ yields the ancilla state
\begin{equation}
\mu_{x}\ket 0+\mu_{y}\ket 1\label{eq: ancilla state (uniform aux)}
\end{equation}
with probability
\begin{align}
p_{xy} & =\frac{1}{2^{n+1}}\left(\left|\mu_{x}\right|^{2}+\left|\mu_{y}\right|^{2}\right).\label{eq: p_xy (uniform aux)}
\end{align}
Eqs. (\ref{eq: ancilla state (uniform aux)}),(\ref{eq: p_xy (uniform aux)})
are the same as $(\ref{eq: ancilla state (RNG)}$),(\ref{eq: p_xy (RNG)}).
Thus, the basic version of the algorithm is equivalent to the general
version with a uniform superposition for $\alpha$. 

\subsection{Derivation of the Variance}

The cost of EVAQS is driven by the number of samples needed to obtain
an estimate with sufficiently small variance. In Appendix \ref{sec: variance derivation}
is is shown that, to lowest order in statistical fluctuations,
\begin{align}
\Var\tilde{F} & \approx\frac{1}{N}\sum_{x}\frac{\left|\tau_{x}\right|^{2}}{\left|\alpha_{x}\right|^{2}}Q_{x}\label{eq: true variance-1}
\end{align}
where
\begin{align}
Q_{x} & \equiv(1+F^{2})\left(\left|\mu_{x}\right|^{2}+\left|\tau_{x}\right|^{2}\right)-2F\left(\tau_{x}^{*}\mu_{x}\braket{\mu}{\tau}+\tau_{x}\mu_{x}^{*}\braket{\tau}{\mu}\right).\label{eq: Q}
\end{align}
In a typical application the expectation is that $\mu$ is not a bad
approximation of $\tau$; thus the regime of interest is that of small
infidelity $I\equiv1-F$. We proceed to simplify $Q_{x}$ for the
case that the infidelity $I\equiv1-F$ is small, keeping only the
lowest order terms. Let $\braket{\tau}{\mu}=e^{\imag\phi}\cos\theta$.
Then $F=\cos^{2}\theta$, $I=\sin^{2}\theta$, and $1+F^{2}\approx2F$.
This yields
\begin{align}
Q_{x} & \approx2F\left(\left|\mu_{x}\right|^{2}+\left|\tau_{x}\right|^{2}-\cos\theta\left(\tau_{x}\mu_{x}^{*}e^{\imag\phi}+\tau_{x}^{*}\mu_{x}e^{-\imag\phi}\right)\right).\label{eq: Q approx 1}
\end{align}
Now, $\mu$ can be written as $\mu=e^{\imag\phi}(\tau\cos\theta+\sigma\sin\theta)$
where $\left\Vert \sigma\right\Vert ^{2}=1$ and $\braket{\sigma}{\tau}=0$.
Then $\tau_{x}^{*}\mu_{x}e^{-\imag\phi}=\left|\tau_{x}\right|^{2}\cos\theta+\tau_{x}^{*}\sigma_{x}\sin\theta$.
To lowest order in $\sin\theta=\sqrt{I}$,
\begin{align}
\left|\mu_{x}\right|^{2} & =\cos^{2}\theta\left|\tau_{x}\right|^{2}+\left|\sigma_{x}\right|^{2}\sin^{2}\theta+\cos\theta\sin\theta\left(\tau_{x}\sigma_{x}^{*}+\tau_{x}^{*}\sigma_{x}\right).
\end{align}
Substituting these expressions into (\ref{eq: Q approx 1}) and combining
terms yields
\begin{align}
Q_{x} & \approx2F\left((1-\cos^{2}\theta)\left|\tau_{x}\right|^{2}+\left|\sigma_{x}\right|^{2}\sin^{2}\theta\right)\\
 & =2I\left(\left|\tau_{x}\right|^{2}+\left|\sigma_{x}\right|^{2}\right).
\end{align}
This yields
\begin{align}
\Var\tilde{F} & \approx\frac{2I}{N}\sum_{x}\frac{\left|\tau_{x}\right|^{2}\left(\left|\tau_{x}\right|^{2}+\left|\sigma_{x}\right|^{2}\right)}{\left|\alpha_{x}\right|^{2}}.\label{eq: variance expression (sigma)}
\end{align}
To obtain eq. (\ref{eq: variance expression (epsilon)}, we combine
the definitions $\mu=e^{\imag\phi}(\tau\cos\theta+\sigma\sin\theta)$
and $\epsilon\equiv e^{-\imag\phi}\mu-\tau$ to obtain
\begin{align}
\epsilon & =\tau(\cos\theta-1)+\sigma\sin\theta.
\end{align}
Since $(I-\tau\tau^{\dagger})\epsilon=\sigma\sin\theta$, $\sigma$
can be understood as the normalized projection of $\epsilon$ onto
the subspace orthogonal to $\tau$. Alternatively, we may use the
fact that $\sin\theta=\sqrt{1-F}=\sqrt{I}$ and $\cos\theta=\sqrt{F}\approx1-\frac{1}{2}I$
to obtain
\begin{align}
\epsilon & \approx\sigma\sqrt{I}.
\end{align}
Eq.\,(\ref{eq: variance expression (epsilon)}) follows from substituting
this expression into (\ref{eq: variance expression (sigma)}).

\subsection{Derivation of the Robustness Bound}

\label{sec: robustness derivation}Suppose $\alpha$ is mischaracterized
as $\tilde{\alpha}$. Then upon measuring $x,y$ one is led to prepare
the ancilla state
\begin{align}
\ket{\tilde{r}_{yx}} & \left(\frac{\tau_{y}}{\tilde{\alpha}_{y}}\ket 0+\frac{\tau_{x}}{\tilde{\alpha}_{x}}\ket 1\right)
\end{align}
and eq. (\ref{eq: measured snippet with ancilla}) becomes
\begin{align}
\ket{\tilde{\psi}_{xy}^{\prime}} & =\frac{1}{\sqrt{2w_{xy}}}\left(\mu_{x}\alpha_{y}\ket 0+\mu_{y}\alpha_{x}\ket 1\right)\left(\frac{\tau_{y}}{\tilde{\alpha}_{y}}\ket 0+\frac{\tau_{x}}{\tilde{\alpha}_{x}}\ket 1\right).
\end{align}
The projection of the ancillas onto the Bell state $\ket{\Phi_{\pm}}$
is
\begin{gather}
\frac{1}{\sqrt{4w_{xy}}}\left(\mu_{x}\tau_{y}\frac{\alpha_{y}}{\tilde{\alpha}_{y}}\pm\mu_{y}\tau_{x}\frac{\alpha_{x}}{\tilde{\alpha}_{x}}\right).
\end{gather}
Comparing this with (\ref{eq: Bell state projection}) shows that
$\tau$ is effectively replaced by $\tilde{\tau}$, where
\begin{align}
\tilde{\tau}_{x} & \equiv\frac{\tau_{x}\alpha_{x}/\tilde{\alpha}_{x}}{\sqrt{\sum_{y}\left|\tau_{y}\alpha_{y}/\tilde{\alpha}_{y}\right|^{2}}}.
\end{align}
That is, error in one's knowledge of the auxiliary state $\alpha$
causes the algorithm to estimate the fidelity of $\mu$ with respect
to a perturbed version of the target state. Intuitively, if $\tilde{\alpha}$
is close to $\alpha$ then $\tilde{\tau}$ will be close to $\tau$
and $F(\mu,\tilde{\tau})$ will be close to $F(\mu,\tau)$. To make
this more precise we use the triangle inequality
\begin{align}
\sqrt{1-F(\mu,\tau)} & \le\sqrt{1-F(\mu,\tilde{\tau})}+\sqrt{1-F(\tau,\tilde{\tau})}
\end{align}
from which it follows that
\begin{align}
F(\mu,\tau) & \ge F(\mu,\tilde{\tau})-2\sqrt{1-F(\tau,\tilde{\tau})}.\label{eq: F error bound (generic)}
\end{align}
In Appendix \ref{sec: robustness bound derivation} it is shown that
\begin{align}
1-F(\tau,\tilde{\tau}) & \le\frac{\delta_{\text{rms}}^{2}}{(1-\delta_{\text{rms}})^{2}}
\end{align}
where $\delta_{\text{rms}}$ is the average relative error of $\tilde{\alpha}$,
defined as
\begin{align}
\delta_{\text{rms}}^{2} & \equiv\sum_{x}\left|\tau_{x}\right|^{2}\left|\frac{\alpha_{x}}{\tilde{\alpha}_{x}}-1\right|^{2}.
\end{align}
 Combining this with (\ref{eq: F error bound (generic)}) yields the
robustness bound (\ref{eq: robustness bound}).

\section{Conclusion}

The verification of complex states produced by quantum computers presents
daunting experimental and computational challenges. In this paper
I presented EVAQS, a novel state verification method that takes a
significant step in addressing these challenges. In this method, a
preparable quantum state is verified against a classical specification
using a combination of relatively simple quantum circuits and on-the-fly
calculations on a conventional computer. In contrast to existing verification
methods, EVAQS is inherently sample-efficient when the target state
is anticoncentrated (i.e., has high entropy) in the chosen measurement
basis. In the case that the target state is not anticoncentrated,
an auxiliary state may be used to importance sample the unknown state,
greatly reducing the number of measurements needed.

The main limitation of EVAQS is the need to calculate selected probability
amplitudes of the target state for comparison to the unknown state.
If the state is not too large (say, less than 30 qubits), the probability
amplitudes may feasibly be calculated ahead of time and stored in
a look-up table for retrieval during verification. In the hopes of
reducing the complexity of classical computation in some cases, the
method was formulated in such a way that the calculated amplitudes
need not be normalized. But if the task is to verify a quantum state
against a classical specification, it seems there is no way to avoid
the need to calculate characteristic features of the target state.
Addressing this computational challenge to quantum state verification
remains an important direction for future work.

EVAQS complements previously known verification methods that are efficient
when the target state is sparse in some readily measurable basis.
However, there exist interesting quantum states that are neither sparse
nor anticoncentrated; for example, coherent analogs of thermal states
at moderate temperatures. Such a state can have an effective support
that is exponentially large (making it challenging for sparse methods)
but still exponentially smaller than the number of basis states (making
it challenging for EVAQS). Other approaches, perhaps yet to be discovered,
will be needed to efficiently verify such states.

\section{Acknowledgments}

This work was performed at Oak Ridge National Laboratory, operated
by UT-Battelle, LLC under contract DE-AC05-00OR22725 for the US Department
of Energy (DOE). Support for the work came from the DOE Advanced Scientific
Computing Research (ASCR) Quantum Testbed Pathfinder Program under
field work proposal ERKJ332. 


\input{verification_of_anticoncentrated_quantum_states_NPJ.bbl}
\appendix

\section{Correction of Residual Bias}

\label{sec: bias correction}To lowest order in statistical fluctuations,
\begin{align}
\left<\frac{\tilde{A}}{\tilde{B}}\right>\approx F & \left(1+\frac{\Cov(\tilde{A},\tilde{B})}{AB}-\frac{\Var\tilde{B}}{B^{2}}\right).
\end{align}
Thus
\[
\frac{\tilde{A}}{\tilde{B}}\left(1-\frac{\Cov(\tilde{A},\tilde{B})}{AB}+\frac{\Var\tilde{B}}{B^{2}}\right)
\]
is an estimator of $F$ in which the lowest order bias ($N^{-1}$)
has been eliminated. The quantities appearing in the correction terms
are unknown, but they can be estimated to the same order of accuracy
as the estimator itself:
\begin{align}
\frac{\Var\tilde{B}}{B^{2}} & \approx\frac{1}{N(N-1)}\sum_{i}\left(\frac{B_{i}}{\tilde{B}}-1\right)^{2}\\
\frac{\Cov(\tilde{A},\tilde{B})}{AB} & \approx\frac{1}{N(N-1)}\sum_{i}\left(\frac{A_{i}}{\tilde{A}}-1\right)\left(\frac{B_{i}}{\tilde{B}}-1\right).
\end{align}
This yields the improved estimator of $F$,
\begin{align}
\tilde{F} & \equiv\frac{\tilde{A}}{\tilde{B}}\left(1+\tilde{C}\right)\label{eq: bias-corrected estimator}
\end{align}
where
\begin{align}
\tilde{C} & =\frac{1}{N(N-1)}\sum_{i}\left(\frac{B_{i}}{\tilde{B}}-\frac{A_{i}}{\tilde{A}}\right)\left(\frac{B_{i}}{\tilde{B}}-1\right).
\end{align}

\section{Evaluation of Some Expectation Values}

\label{sec: variance derivation}The variance of $\tilde{F}$ is a
function of several different expectation values, which are here identified
and evaluated. To lowest order in statistical fluctuations, the variance
of $\tilde{F}$ is
\begin{align}
\Var\tilde{F} & =\Var\frac{\tilde{A}}{\tilde{B}}\\
 & \approx\frac{\langle\tilde{A}\rangle^{2}}{\langle\tilde{B}\rangle^{2}}\Var\left(\frac{\tilde{A}}{\langle\tilde{A}\rangle}-\frac{\tilde{B}}{\langle\tilde{B}\rangle}\right).
\end{align}
With the relations $\left<\tilde{A}\right>=\lambda F$ and $\left<\tilde{B}\right>=\lambda$
this simplifies to
\begin{align}
\Var\tilde{F} & \approx F^{2}\Var\left(\frac{\tilde{A}}{\lambda F}-\frac{\tilde{B}}{\lambda}\right)\\
 & =\Var\left(\frac{1}{\lambda}\left(\tilde{A}-\tilde{B}F\right)\right)\\
 & =\frac{1}{\lambda^{2}}\left(\left<\left(\tilde{A}-\tilde{B}F\right)^{2}\right>-\left<\tilde{A}-\tilde{B}F\right>^{2}\right).
\end{align}
Now, $\left<\tilde{A}-\tilde{B}F\right>=0$. To evaluate $\left<\left(\tilde{A}-\tilde{B}F\right)^{2}\right>$
we use (\ref{eq: A estimator}) and (\ref{eq: B estimator}) to write
\begin{align}
\tilde{A}-\tilde{B}F & =\frac{\lambda}{N}\sum_{i=1}^{N}w_{x_{i}.y_{i}}\left(b_{i}-b_{i}^{2}F\right)\\
 & \equiv\frac{\lambda}{N}\sum_{i=1}^{N}G_{i}
\end{align}
where each $G_{i}=w_{x_{i}.y_{i}}\left(b_{i}-b_{i}^{2}F\right)$ is
an independent random variable with mean 0. Then
\begin{align}
\Var\tilde{F} & \approx\frac{1}{N^{2}}\sum_{i,j=1}^{N}\left<G_{i}G_{j}\right>\\
 & =\frac{1}{N^{2}}\sum_{i=1}^{N}\left<G_{i}^{2}\right>\\
 & =\frac{1}{N}\left<w_{xy}^{2}(b-b^{2}F)^{2}\right>.
\end{align}
Using $b^{3}=b$ and $b^{4}=b^{2}$ we obtain
\begin{align}
\Var\tilde{F} & \approx\frac{1+F^{2}}{N}\left<w_{xy}^{2}b^{2}\right>-\frac{2F}{N}\left<w_{xy}^{2}b\right>.
\end{align}
Now,
\begin{align}
\left<w_{xy}^{2}b^{2}\right> & =\sum_{x,y,b}w_{xy}^{2}p_{xyb}b^{2}\\
 & =\frac{1}{2}\sum_{x,y}\left(\frac{\left|\tau_{x}\right|^{2}}{\left|\alpha_{x}\right|^{2}}+\frac{\left|\tau_{y}\right|^{2}}{\left|\alpha_{y}\right|^{2}}\right)\left(\left|\mu_{x}\right|^{2}\left|\tau_{y}\right|^{2}+\left|\mu_{y}\right|^{2}\left|\tau_{x}\right|^{2}\right)\\
 & =\sum_{x}\frac{\left|\tau_{x}\right|^{2}}{\left|\alpha_{x}\right|^{2}}\left(\left|\mu_{x}\right|^{2}+\left|\tau_{x}\right|^{2}\right)
\end{align}
and
\begin{align}
\left<w_{xy}^{2}b\right> & =\sum_{x,y,b}w_{xy}^{2}p_{xyb}b\\
 & =\frac{1}{2}\sum_{x,y}\left(\frac{\left|\tau_{x}\right|^{2}}{\left|\alpha_{x}\right|^{2}}+\frac{\left|\tau_{y}\right|^{2}}{\left|\alpha_{y}\right|^{2}}\right)\left(\mu_{x}\tau_{y}\mu_{y}^{*}\tau_{x}^{*}+\mu_{y}\tau_{x}\mu_{x}^{*}\tau_{y}^{*}\right)\\
 & =\sum_{x}\frac{\left|\tau_{x}\right|^{2}}{\left|\alpha_{x}\right|^{2}}\left(\tau_{x}^{*}\mu_{x}\braket{\mu}{\tau}+\tau_{x}\mu_{x}^{*}\braket{\tau}{\mu}\right).
\end{align}
This gives
\begin{align}
\Var\tilde{F} & \approx\frac{1}{N}\sum_{x}\frac{\left|\tau_{x}\right|^{2}}{\left|\alpha_{x}\right|^{2}}Q_{x}
\end{align}
where $Q_{x}$ is given by eq.~(\ref{eq: Q}).

\section{A Bound for the Robustness Result}

\label{sec: robustness bound derivation}In section \ref{sec: robustness}
it was shown that error in the auxiliary state $\alpha$ translates
to an effective error in the target state $\tau$. In this section
I bound the impact of such error on the estimated fidelity. Let $p$
denote the probability distribution induced by $\tau$, i.e. $p_{x}=\left|\tau_{x}\right|^{2}$.
Let $\delta$, defined via $\alpha_{x}/\tilde{\alpha}_{x}=1+\delta_{x}$,
quantify the error in the auxiliary state. In terms of these quantities
$F(\tau,\tilde{\tau})$ can be written as
\begin{align}
F(\tau,\tilde{\tau}) & =\frac{\left|\sum_{x}p_{x}(1+\delta_{x})\right|^{2}}{\sum_{x}p_{x}\left|1+\delta_{x}\right|^{2}}\\
 & =\frac{\left|\left<1+\delta_{x}\right>\right|^{2}}{\left<\left|1+\delta_{x}\right|^{2}\right>}
\end{align}
where expectations are taken with respect to $p$. Then
\begin{align}
1-F(\tau,\tilde{\tau}) & =1-\frac{\left|\left<1+\delta_{x}\right>\right|^{2}}{\left<\left|1+\delta_{x}\right|^{2}\right>}\\
 & =\frac{\left<\left|1+\delta_{x}\right|^{2}\right>-\left|\left<1+\delta_{x}\right>\right|^{2}}{\left<\left|1+\delta_{x}\right|^{2}\right>}\\
 & =\frac{\Var\left(1+\delta_{x}\right)}{\left<\left|1+\delta\right|^{2}\right>}.
\end{align}
An upper bound on the numerator is
\begin{align}
\Var\left(1+\delta_{x}\right) & =\Var\delta_{x}\le\delta_{\text{rms}}^{2}
\end{align}
where $\delta_{\text{rms}}\equiv\left<\left|\delta_{x}\right|^{2}\right>^{\frac{1}{2}}$.
For the denominator, non-negativity of variance gives
\begin{align}
\left<\left|1+\delta_{x}\right|^{2}\right> & \ge\left|\left<1+\delta_{x}\right>\right|^{2}.
\end{align}
Now, $\left|1+\left<\delta_{x}\right>\right|\ge1-\left|\left<\delta_{x}\right>\right|$.
This time non-negativity of variance gives $\left|\left<\delta_{x}\right>\right|\le\delta_{\text{rms}}$,
hence
\begin{align}
\left|1+\left<\delta_{x}\right>\right| & \ge1-\delta_{\text{rms}}.
\end{align}
The upper bound on the numerator combined with the lower bound on
the denominator yields
\begin{align}
1-F(\tau,\tilde{\tau}) & \le\frac{\delta_{\text{rms}}^{2}}{(1-\delta_{\text{rms}})^{2}}.
\end{align}
 
\end{document}

%% file: macros.tex
\global\long\def\sinc{\operatorname{sinc}}%

\global\long\def\rank{\operatorname{rank}}%

\global\long\def\vspan{\operatorname{span}}%

\global\long\def\Tr{\operatorname{Tr}}%

\global\long\def\Re{\operatorname{Re}}%

\global\long\def\Im{\operatorname{Im}}%

\global\long\def\ket#1{|#1\rangle}%

\global\long\def\bra#1{\langle#1|}%

\global\long\def\braket#1#2{\langle#1|#2\rangle}%

\global\long\def\Ket#1{\left|#1\right\rangle }%

\global\long\def\Bra#1{\left\langle #1\right|}%

\global\long\def\cket#1{|#1]}%

\global\long\def\cbra#1{[#1|}%

\global\long\def\imag{\mathrm{i}}%

\global\long\def\deriv{\mathrm{d}}%

{\LARGE{}}
\global\long\def\op#1{\operatorname\{#1\}}%
{\LARGE\par}

{\LARGE{}}
\global\long\def\Var{\operatorname{Var}}%
{\LARGE\par}

{\LARGE{}}
\global\long\def\Vol{\operatorname{Vol}}%
{\LARGE\par}

{\LARGE{}}
\global\long\def\Cov{\operatorname{Cov}}%
{\LARGE\par}

{\LARGE{}}
\global\long\def\sign{\operatorname{sign}}%
{\LARGE\par}

{\LARGE{}}
\global\long\def\eig{\operatorname{eig}}%
{\LARGE\par}

{\LARGE{}}
\global\long\def\diag{\operatorname{diag}}%
{\LARGE\par}

{\LARGE{}}
\global\long\def\Nb{\operatorname{Nb}}%
{\LARGE\par}

{\LARGE{}}
\global\long\def\Perm{\operatorname{Perm}}%
{\LARGE\par}

{\LARGE{}}
\global\long\def\perm{\operatorname{perm}}%
{\LARGE\par}